\RequirePackage{fix-cm}
\documentclass[a4paper,11pt]{article}

\usepackage{authblk}
\usepackage{amsfonts,amsmath,amssymb,amsthm}
\usepackage{algorithm}
\usepackage{paralist}
\usepackage{booktabs}
\usepackage{url}
\usepackage{hyperref}
\usepackage{graphicx}
\usepackage{microtype}
\usepackage{url}
\usepackage{hyperref}
\usepackage{subcaption}
\usepackage{tikz}
\usepackage{multirow}
\usetikzlibrary{positioning}
\usetikzlibrary{decorations.pathreplacing}
\newcommand{\N}{\mathbb{N}}

\newcommand{\F}{\mathbb{F}}

\DeclareMathOperator{\rank}{rank}
\DeclareMathOperator{\nullity}{null}
\newcommand{\cat}{\mathbin\Vert}

\newtheorem{definition}{Definition}
\newtheorem{lemma}{Lemma}
\newtheorem{theorem}{Theorem}
\newtheorem{remark}{Remark}

\tikzset{
    mybrace/.style={decorate,decoration={brace,aspect=#1}}
}

\providecommand{\keywords}[1]{\textbf{\textit{Keywords }} #1}

\begin{document}

\title{On the Minimum Distance of Subspace Codes Generated by Linear Cellular Automata}

\author[1]{Luca Mariot}
\author[1]{Federico Mazzone}
	
\affil[1]{{\small Semantics, Cybersecurity and Services Group, University of Twente, Drienerlolaan 5, 7511GG Enschede, The Netherlands} 
	
	{\small \texttt{\{l.mariot, f.mazzone\}@utwente.nl}}}

\maketitle

\begin{abstract}
Motivated by applications to noncoherent network coding, we study subspace codes defined by sets of linear cellular automata (CA). As a first remark, we show that a family of linear CA where the local rules have the same diameter---and thus the associated polynomials have the same degree---induces a Grassmannian code. Then, we prove that the minimum distance of such a code is determined by the maximum degree occurring among the pairwise greatest common divisors (GCD) of the polynomials in the family. Finally, we consider the setting where all such polynomials have the same GCD, and determine the cardinality of the corresponding Grassmannian code. As a particular case, we show that if all polynomials in the family are pairwise coprime, the resulting Grassmannian code has the highest minimum distance possible.
\end{abstract}


\section{Introduction}
\label{sec:intro}
The conventional way of routing packets from source to sink nodes frequently fails to exploit a network's full potential, which is a common issue in networking. The \emph{butterfly network}~\cite{kschischang12} serves as a classic example of this problem. The field of network coding emerged around two decades ago, and seeks to solve this problem by exploiting a simple idea: instead of simply routing packets, intermediate nodes in the network can \emph{combine} them, usually by employing linear operators~\cite{medard2011}. In this way, more packets can be multiplexed over a single channel usage. In the \emph{noncoherent} network coding strategy, the messages transmitted between nodes are subspaces of an ambient vector space~\cite{koetter08}. In this scenario, the need to encode and decode subspaces in a reliable way for transmission over networks spawned a branch of coding theory that deals with \emph{subspace codes}~\cite{khaleghi09}. These codes can be seen as a generalization of classic linear error correcting codes, where the codewords are subspaces rather than vectors. By embedding the projective space of a vector space with a suitable metric, it is possible to define the minimum distance between any two codewords in a subspace code. Similarly to the usual case of linear error correcting codes, it is desirable to define codes containing a large number of subspaces (to maximize the network's capacity) such that they are at the highest possible distance from each other (to correct as many errors and erasures as possible).

The aim of this paper is to explore the idea of using cellular automata (CA) to construct subspace codes. We consider the specific case of linear CA, motivated by the fact that the body of literature concerning them is quite extensive. Previous work~\cite{mariot20} focused on a construction of maximal sets of mutually orthogonal Latin squares (MOLS) based on linear bipermutive CA. Such a construction is equivalent to finding a maximal family of pairwise coprime polynomials over a finite field, all having the same degree and a nonzero constant term.
Here, we investigate another research question stemming from this construction: \emph{what kind of subspace codes can be obtained by families of linear CA, if the underlying polynomials that define their local rules are not pairwise coprime?}

The main contributions of this work are listed below:
\begin{compactitem}
\item We show that a family of linear CA with local rules of the same diameter generates a \emph{constant dimension code}, also known as a Grassmannian code~\cite{Etzion19}.
\item We characterize the minimum distance of a Grassmannian code generated by a family of linear CA.
\item We observe that the minimum distance of such codes is optimal when the defining polynomials are pairwise coprime. This is the case considered for MOLS and bent functions~\cite{gadouleau20,gadouleau23}, with the resulting Grassmannian codes being a particular breed of the partial spread codes introduced in~\cite{gorla14}.
\item We study the specific case where the rules of a family of linear CA are defined by polynomials that have the same GCD, and determine the number of codewords in the resulting Grassmannian code.
\end{compactitem}

The remainder of this paper is organized as follows. Section~\ref{sec:basic} recalls all background notions related to cellular automata and subspace codes that are necessary to introduce our results. Section~\ref{sec:sc-ca} formally defines the subspace code generated by a family of linear CA, and remarks that if the underlying local rules have all the same diameter, the resulting code has constant dimension. Section~\ref{sec:min-dist} proves the main result of the paper, namely the relationship between the minimum distance of a Grassmannian code generated by linear CA and the maximum degree occurring among the pairwise GCDs of the associated polynomials. Section~\ref{sec:cdc-ca} analyzes the cardinality of the Grassmannian codes induced by linear CA whose underlying polynomials have the same pairwise GCD. Finally, Section~\ref{sec:outro} summarizes the key contributions of the paper, and elaborates on several directions and open problems for future research on the topic.

\section{Basic Definitions}
\label{sec:basic}
In this section, we cover all the basic definitions and results related to cellular automata and subspace codes used throughout the paper. As a general notation, given $q \in \N$ a power of a prime, we denote by $\F_q$ the finite field of order $q$. For all $n \in \N$, the set of all $n$-tuples over $\F_q$ is denoted by $\F_q^n$, and we endow it with the structure of a vector space, where vector sum and multiplication by a scalar are inherited in the usual way from the sum and product operations of $\F_q$.

\subsection{Cellular Automata}
\label{subsec:ca}
A Cellular Automaton (CA) is a type of discrete dynamical system that consists of a regular lattice of cells, which can be either finite or infinite. Each cell updates its state based on a local rule that is applied to its own state and the states of its neighboring cells. This updating process occurs simultaneously for all cells in the lattice, and it is repeated over multiple time steps, giving rise to the dynamic behavior of the system. If the lattice is finite, periodic boundary conditions are typically assumed. This ensures that each cell always has enough neighbors to evaluate the local rule.

While most research on CA focuses on their long-term dynamical behavior, in this work we consider CA as algebraic systems. Specifically, the local rule is applied only once, and only by cells that have enough neighbors to evaluate it. This leads to a CA model that can be viewed as a particular type of vectorial functions over finite fields, which we formally define below:

\begin{definition}
\label{def:ca}
Let $d,n \in \N$ such that $d \le n$, and set $k=d-1$. Further, let $f: \F_q^d \to \F_q$ be a $d$-variable function over the finite field $\F_q$. A \emph{cellular automaton} of length $n$, diameter $d$, and local rule $f$ is a vectorial mapping $F: \F_q^{n} \to \F_q^{n-k}$ whose $i$-th output coordinate is defined as:
\begin{equation}
\label{eq:ca}
F(x_0,\cdots,x_{n-1})_i = f(x_i,\cdots, x_{i+k})
\end{equation}
for all $i \in \{0,\cdots,n-k-1\}$ and $x \in \F_q^n$.
\end{definition}
Intuitively, the output coordinate $F_i$ consists in the application of the local rule $f$ over the neighborhood formed by the $i$-th input coordinate and the $k$ coordinates on its right. This is the reason why the function maps the vector space $\F_q^n$ to the smaller subspace $\F_q^{n-k}$: the local rule is applied as long as we have enough right neighbors, i.e., up to the $(n-k)$-th coordinate. Thus, the cellular lattice size is reduced by $k$ coordinates after evaluating $F$. As we mentioned above, this is not a problem, since we consider only the one-shot application of $F$ and we are not interested in iterating the CA over multiple time steps.

In this work we focus on \emph{linear} CA, where the local rule is a linear combination of the input coordinates, that is, for all $x \in \F_q^d$ we have: 
\begin{equation}
\label{eq:lin-f}
f(x_0,\cdots, x_{k}) = a_0x_0 + a_1x_1 + \cdots + a_kx_k
\end{equation}
for some $a_0, \dots, a_k \in \F_q$.
Further, one can associate to each linear rule of the form \eqref{eq:lin-f} a polynomial $P_f \in \F_q[X]$ in a natural way as follows:
\begin{equation}
\label{eq:pol-f}
P_f(X) = a_0 + a_1X + \cdots + a_kX^k \enspace .
\end{equation}
In other words, we use the coefficients of the vector $(a_0,\cdots,a_k)$ that define the local rule as the coefficients of the monomials $X^i$, in increasing order of powers. In what follows, we will assume that $a_0,a_k \neq 0$, and in particular that $a_k=1$. This implies that the local rule is \emph{bipermutive}, since any restriction of $f$ obtained by fixing either the first or the last $k$ input variables induces a permutation of $\F_q$ respectively on the last or on the first variable~\cite{mariot20}. Moreover, the polynomial associated to $f$ is monic of degree $k$ and has a nonzero constant term.

\subsection{Subspace Codes}
\label{subsec:sc}
In this section we cover only the basic notions related to subspace codes. We refer the reader to~\cite{koetter08} for a more comprehensive treatment of the subject.

We start by considering the vector space $\F_q^n$. We denote by $\mathcal{P}(\F_q^n)$ its \emph{projective space}, i.e., the family of all subspaces of $\F_q^n$. Usually, in the context of network coding, the projective space is interpreted as a metric space under the following distance: for all $A,B \in \mathcal{P}(\F_q^n)$, we have
\begin{equation}
    \label{eq:dist}
    d(A,B) = dim(A) + dim(B) - 2dim(A \cap B) \enspace .
\end{equation}
We can now introduce the definition of subspace code:
\begin{definition}
\label{def:sc}
Let $n \in \N$. A subspace code $\mathcal{C}$ of parameters $[n, \ell(C), \log_q|\mathcal{C}|, D(\mathcal{C})]$ is a subset of $\mathcal{P}(\F_q^n)$ where $\ell(\mathcal{C}) = \max_{V \in \mathcal{C}} \{dim(V)\}$ and $D(\mathcal{C})$ is the minimum distance of $\mathcal{C}$, defined as:
\begin{equation}
    \label{eq:mindist}
    D(\mathcal{C}) = \min_{U,V \in \mathcal{C}} \left\{d(U,V)\right\} \enspace ,
\end{equation}
where $d(\cdot,\cdot)$ is computed as in~\eqref{eq:dist}.
\end{definition}
This definition generalizes the concept of error-correcting codes by considering codewords that are subspaces rather than vectors. In other words, the elements of the code are not individual vectors, but sets of vectors that span a subspace of the underlying vector space.

The set of all subspaces of dimension $k$, for a given $0 \le k \le n$, is also called the \emph{Grassmannian}, and it is denoted by $Gr(\F_q^n, k)$. Accordingly, a subspace code $\mathcal{C} \subseteq Gr(\F_q^n, k)$ is known as a \emph{Grassmannian code}, or equivalently a \emph{constant dimension code}, since each subspace in $\mathcal{C}$ has dimension $k$, i.e. $\ell(\mathcal{C}) = k$.

The main problem studied for subspace codes is analogous to the one studied for classic error-correcting codes: for a fixed minimum distance $\delta$, what is the maximum cardinality achievable by a subspace code $\mathcal{C}$ with $D(\mathcal{C})=\delta$? Intuitively, the lower the allowed minimum distance $\delta$ is, the more subspaces we can pack together in a code---and therefore, in the context of network coding, the more messages we can transmit over a network. On the other hand, one also wants that any two subspaces are as far as possible from each other for error-correction purposes, or equivalently a subspace code with the highest minimum distance possible. In the following sections we explore this trade-off for subspace codes defined by families of linear CA.

\section{Subspaces Codes from Families of Linear CA}
\label{sec:sc-ca}
We now describe our method to construct subspace codes using sets of linear CA. Suppose that $F: \F_q^n \to \F_q^{n-k}$ is a linear CA defined by a local rule $f: \F_q^d \to \F_q$ of diameter $d$ with associated polynomial $P_f(X) = a_0+a_1X+\cdots+a_kX^k$ where $k=d-1$. This CA is a linear mapping of the form $F(x) = M_F\cdot x^\top$ for all $x \in \F_q^n$, where $M_F$ is a $k\times n$ matrix over $\F_q$ of the following form:
\begin{equation}
\label{eq:trans-mat}
M_F =
\begin{pmatrix}
a_0 & \dots & a_k & 0 & 0 & \dots & 0 \\
0 & a_0 & \dots & a_k & 0 & \dots & 0 \\
\vdots & \cdots & \ddots & \ddots & \ddots & \cdots & \vdots \\
0 & \dots & 0 & 0 & a_0 & \dots & a_k \\
\end{pmatrix} \enspace .
\end{equation}
The matrix $M_F$ is called the \textit{transition matrix} of $F$, and it is obtained by shifting the coefficients of $P_f$ one place to the right per each subsequent row. As we discussed in Section~\ref{subsec:ca}, we assume that $a_0\neq 0$ and $a_k=1$. In this way, the polynomial $P_f$ is monic of degree $k$ with a nonzero constant term, and all the columns of $M_F$ are nonzero.

Now, let us consider the \emph{kernel} $ker(f)$ of the linear CA $F$. By definition, this is the subspace of input vectors $x \in \F_q^n$ such that $F(x) = \underbar{0}$, and it is equivalent to the nullspace of the matrix $M_F$. A method to construct the kernel of $F$ is by using the following \emph{preimage computation} procedure:
\begin{compactitem}
\item Set the output configuration of the CA $F$ to the null vector $\underbar{0}$.
\item For all $\tilde{x} = (\tilde{x}_0,\tilde{x}_1,\cdots,\tilde{x}_{k-1}) \in \F_q^k$, do:
\begin{compactenum}
\item Set the first $k$ coordinates of the CA input $x \in \F_q^n$ to $\tilde{x}$, that is, $x_0 = \tilde{x}_0$, $x_1 = \tilde{x}_1$, $\cdots$, $x_{k-1} = \tilde{x}_{k-1}$.
\item For all $i \in \{k,\cdots,n-1\}$, compute the $i$-th input coordinate $x_i$ as:
\begin{equation}
\label{eq:inv-bip}
x_i = -(a_0x_{i-k} + a_1x_{i-k+1} + \cdots + a_{k-1}x_{i-1}) \enspace .
\end{equation}
\end{compactenum}
\end{compactitem}
Equation~\eqref{eq:inv-bip} stems from the fact that the $i$-th output coordinate of the CA must be zero (since the whole output configuration is the null vector $\underbar{0}$). Thus, one can recover $x_i$ from the equation of the local rule, moving all the terms $a_0x_{i-k}, \cdots, a_{k-1}x_{i-1}$ to the left hand side and changing their sign. Since we assume $a_k=1$, one then obtains Equation~\eqref{eq:inv-bip}.

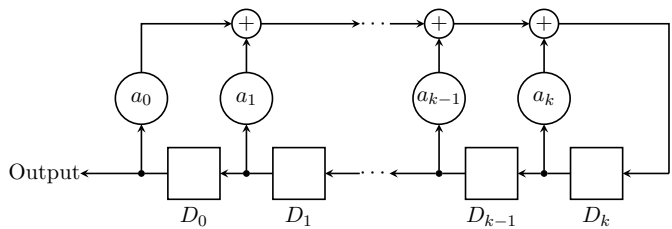
\begin{figure}[t]
\centering
\resizebox{!}{3cm}{
\begin{tikzpicture}
[->,auto,node distance=1cm,
       empt node/.style={minimum width=4pt, minimum height=2pt,inner
         sep=0pt,outer sep=0pt},
       empt2 node/.style={minimum width=0pt, minimum height=0pt,inner sep=0pt,outer sep=0pt},
       rect node/.style={rectangle,thick,draw,font=\sffamily\bfseries,minimum
         width=0.9cm,minimum height=0.9cm, inner sep=0pt, outer sep=0pt},
       circ node/.style={circle,thick,draw,font=\sffamily\bfseries,minimum
         width=0.9cm, inner sep=0pt, outer sep=0pt},
       circ2 node/.style={circle,thick,draw,font=\sffamily\bfseries,minimum
         width=0.5cm, inner sep=0pt, outer sep=0pt},
       dot node/.style={circle,draw,fill=black,inner sep=1pt}]

\node [rect node, label={below:$D_0$}] (d0) {};
\node [dot node] (p0) [left=0.4cm of d0] {};
\node [empt node] (out) [left=1cm of p0] {Output};
\node [circ node] (a0) [above=0.8cm of p0] {$a_0$};
\node [dot node] (p1) [right=0.4cm of d0] {};
\node [circ node] (a1) [above=0.8cm of p1] {$a_1$};
\node [circ2 node] (s1) [above=0.6cm of a1] {$+$};
\node [rect node, label={below:$D_1$}] (d1) [right=0.4cm of p1] {};
\node [empt node] (ddots) [right=0.6cm of d1] {$\cdots$};
\node [dot node] (p2) [right=0.8cm of ddots] {};
\node [circ node] (ak1) [above=0.8cm of p2] {$a_{k-1}$};
\node [circ2 node] (sk1) [above=0.6cm of ak1] {$+$};
\node [empt node] (sdots) [left=0.6cm of sk1] {$\cdots$};
\node [rect node, label={below:$D_{k-1}$}] (dk1) [right=0.4cm of p2] {};
\node [dot node] (p3) [right=0.4cm of dk1] {};
\node [circ node] (ak) [above=0.8cm of p3] {$a_{k}$};
\node [circ2 node] (sk) [above=0.6cm of ak] {$+$};
\node [rect node, label={below:$D_{k}$}] (dk) [right=0.4cm of p3] {};
\node [empt2 node] (con) [right=0.8cm of dk] {};

\draw [->, thick, shorten >=0pt,shorten <=0pt,>=stealth] (p0) -- (out);
\draw [->, thick, shorten >=0pt,shorten <=0pt,>=stealth] (p0) -- (a0);
\draw [->, thick, shorten >=0pt,shorten <=0pt,>=stealth] (a0) |- (s1);
\draw [-, thick, shorten >=0pt,shorten <=0pt,>=stealth] (d0) -- (p0);
\draw [->, thick, shorten >=0pt,shorten <=0pt,>=stealth] (p1) -- (d0);
\draw [->, thick, shorten >=0pt,shorten <=0pt,>=stealth] (p1) -- (a1);
\draw [->, thick, shorten >=0pt,shorten <=0pt,>=stealth] (a1) -- (s1);
\draw [->, thick, shorten >=0pt,shorten <=0pt,>=stealth] (s1) -- (sdots);
\draw [->, thick, shorten >=0pt,shorten <=0pt,>=stealth] (sdots) -- (sk1);
\draw [->, thick, shorten >=0pt,shorten <=0pt,>=stealth] (sk1) -- (sk);
\draw [-, thick, shorten >=0pt,shorten <=0pt,>=stealth] (d1) -- (p1);
\draw [->, thick, shorten >=0pt,shorten <=0pt,>=stealth] (ddots) -- (d1);
\draw [->, thick, shorten >=0pt,shorten <=0pt,>=stealth] (p2) -- (ddots);
\draw [->, thick, shorten >=0pt,shorten <=0pt,>=stealth] (p2) -- (ak1);
\draw [-, thick, shorten >=0pt,shorten <=0pt,>=stealth] (dk1) -- (p2);
\draw [->, thick, shorten >=0pt,shorten <=0pt,>=stealth] (p3) -- (dk1);
\draw [->, thick, shorten >=0pt,shorten <=0pt,>=stealth] (p3) -- (ak);
\draw [->, thick, shorten >=0pt,shorten <=0pt,>=stealth] (ak1) -- (sk1);
\draw [->, thick, shorten >=0pt,shorten <=0pt,>=stealth] (ak) -- (sk);
\draw [-, thick, shorten >=0pt,shorten <=0pt,>=stealth] (dk) -- (p3);
\draw [-, thick, shorten >=0pt,shorten <=0pt,>=stealth] (sk) -| (con);
\draw [->, thick, shorten >=0pt,shorten <=0pt,>=stealth] (con) -- (dk);
       
\end{tikzpicture}
}
\caption{Example of linear feedback shift register of order $k$.}
\label{fig:lfsr}
\end{figure}

This preimage computation procedure is equivalent to the computation of a $k$-th order homogeneous \emph{Linear Recurring Sequence} (LRS)~\cite{lidl1997finite}. In particular, the kernel of $F$ corresponds to all infinite sequences $x_0, x_1,\cdots$ of elements in $\F_q$ that satisfy the following recurrence equation:
\begin{equation}
\label{eq:lrs}
a_0x_i + a_1x_{i+1} + \cdots + a_kx_{i+k} = 0 \enspace ,
\end{equation}
and truncating such sequences to the $n$-th element. Thus, one can compute the kernel of $F$ by using a \emph{Linear Feedback Shift Register} (LFSR) of order $k$ and with feedback polynomial $P_f$ (see Figure~\ref{fig:lfsr}).
The idea is to initialize the registers $D_0,\cdots,D_{k-1}$ with the starting block $\tilde{x} \in \F_q^k$ of the preimage $x$, and then run the LFSR for $n$ clock steps. At each step $i \in \{0,\cdots, n-1\}$, the rightmost register $D_k$ is updated with the feedback of the linear recurrence equation~\eqref{eq:lrs}, while the leftmost register $D_0$ outputs the value of $x_i$. We remark that this approach has been adopted in~\cite{mariot17} to study the period of spatially periodic preimages in linear bipermutive CA and in~\cite{mariot18} to construct cyclic codes from linear CA.

From the discussion above, we can conclude the following result:
\begin{lemma}
\label{lm:dim-ker}
Let $F: \F_q^n \to \F_q^{n-k}$ be a linear $CA$ defined by a local rule $f:\F_q^d \to \F_q$ of diameter $d$, where $k=d-1$, and denote by $M_F$ the transition matrix of $F$. Then, $dim(ker(F)) = k$ and $\rank(M_F) = n-k$.
\begin{proof}
Using the preimage computation procedure outlined above, the number of preimages of the null vector $\underbar{0}$ under $F$ is $q^k$, since each of them is uniquely determined by a vector $\tilde{x} \in \F_q^k$. Hence, $|ker(F)| = q^k$, and $dim(ker(F)) = k$. The rank of $M_F$ now follows from the fact that $ker(F)$ is also the nullspace of $M_F$, and from the rank-nullity theorem: the number of columns of $M_F$ equals the sum of the rank of $M_F$ and the dimension of its nullspace. \qed
\end{proof}
\end{lemma}

We are now ready to define a subspace code generated by a set of linear CA.
\begin{definition}
\label{def:sc-ca}
Let $n,d \in \N$ with $d\le n$ and $k=d-1$. The subspace code generated by a family $\mathcal{F}$ of $t$ linear CA $F_1,\cdots, F_t: \F_q^n \to \F_q^{n-k}$, respectively defined by bipermutive local rules $f_1,\cdots,f_t: \F_q^d \to \F_q$ of diameter $d$, is the set
\begin{equation}
\label{eq:sc-ca}
\mathcal{C}_{\mathcal{F}} = \{ ker(F_i): 1\le i \le t \} \enspace .
\end{equation}
\end{definition}
In other words, the subspace code consists of the kernels of all the $t$ linear CA in the family $\mathcal{F}$. The reader might wonder why we choose specifically the kernels of the CA instead of, for instance, their images. This will become clearer in the next section where we exploit this fact to characterize the minimum distance of the code. Moreover, from Lemma~\ref{lm:dim-ker} it holds that each kernel in $\mathcal{C}_\mathcal{F}$ has dimension $k$. Thus, we have the following result:
\begin{lemma}
\label{lm:cdc}
The subspace code $\mathcal{C}_\mathcal{F}$ defined in Equation~\eqref{eq:sc-ca} is a Grassmannian code, i.e. $\mathcal{C}_\mathcal{F} \subseteq Gr(\F_q^n, k)$.
\end{lemma}

\section{Relation between Minimum Distance and GCD}
\label{sec:min-dist}
Lemma~\ref{lm:cdc} prompts us with the following natural question: is it possible to characterize the minimum distance of a Grassmannian code generated by a family $\mathcal{F}$ of linear CA, possibly linking it with the properties of the polynomials associated to the local rules? In this section, we analyze this issue.

In the following discussion, we make the assumption that $n=2k$. Hence, a subspace code is generated by a family of linear CA $F_1,\cdots,F_t: \F_q^{2k}\to \F_q^{k}$.
The codewords of the Grassmannian code $\mathcal{C}_\mathcal{F}$ are the kernels $ker(F_i)$ for $1\le i \le t$. By applying~\eqref{eq:dist}, and Lemma~\ref{lm:cdc}, the distance between any two kernels in $\mathcal{C}_\mathcal{F}$ equals:
\begin{align}
\label{eq:dist-ker}
\nonumber
d(ker(F),ker(G)) &= dim(ker(F)) + dim(ker(G)) - 2dim(ker(F) \cap ker(G)) = \\
                   &= 2k - 2dim(ker(F) \cap ker(G)) \enspace .
\end{align}
Thus, this distance is inversely proportional to the size of the intersection of the kernels. We can then characterize the minimum distance $D(\mathcal{C}_F)$ in terms of the largest intersection between any two kernels in the subspace code. To this end, we first need some further results. Given any two CA $F,G \in \mathcal{F}$, with local rules $f,g$ respectively, we can define their concatenation $H: \F_q^{2k} \to \F_q^{2k}$ as the map
\begin{equation}
H(x) := F(x) \cat G(x) \enspace .
\end{equation}

\begin{remark}
\label{rem:int-ker}
We can easily see that $H$ is still a linear application, and $H(x) = \underbar{0}$ if and only if $F(x) = \underbar{0}$ and $G(x) = \underbar{0}$. So we have that the kernel of $H$ is nothing else than the intersection $ker(F) \cap ker(G)$.
\end{remark}

The matrix associated to $H$ is the vertical concatenation of $M_F$ and $M_G$:
\begin{equation}
\label{eq:sylv}
M_H =
\begin{pmatrix}
a_0 & \dots & a_k & 0 & 0 & \dots & 0 \\
0 & a_0 & \dots & a_k & 0 & \dots & 0 \\
\vdots & \cdots & \ddots & \ddots & \ddots & \cdots & \vdots \\
0 & \dots & 0 & 0 & a_0 & \dots & a_k \\
b_0 & \dots & b_k & 0 & 0 & \dots & 0 \\
0 & b_0 & \dots & b_k & 0 & \dots & 0 \\
\vdots & \cdots & \ddots & \ddots & \ddots & \cdots & \vdots \\
0 & \dots & 0 & 0 & b_0 & \dots & b_k \\
\end{pmatrix} \enspace .
\end{equation}
We can recognize such matrix as the \emph{Sylvester matrix }associated to the polynomials $P_f, P_g$ corresponding to the local rules $f,g$. Notably, the determinant of this matrix is called the \emph{resultant} of $P_f$ and $P_g$, denoted by $Res(P_f,P_g)$, and it is known that $Res(P_f,P_g) \neq 0 \Leftrightarrow \gcd(P_f,P_g) = 1$~\cite{gelfand}. In other words, the Sylvester matrix $M_H$ is invertible if and only if the two polynomials $P_f, P_g$ defining the local rules of $F$ and $G$ are relatively prime. This fact was used by the authors of~\cite{mariot20} to construct orthogonal Latin squares from linear CA.

In our setting of Grassmannian codes, we are interested in the more general situation where the Sylvester matrix associated to $F$ and $G$ is not necessarily invertible. To determine the dimension of the intersection of $ker(F)$ and $ker(G)$ we need the following result that links the dimension of the null space of the Sylvester matrix to the degree of the GCD of the two polynomials\footnote{This seems to be a widely known result, but we could not find any reference in the literature that proves it. Hence, we report a proof here for our convenience.}:

\begin{lemma}
\label{lm:rank-syl}
Let $f,g \in \F_q[X]$ be two polynomials, and denote by $S_{f,g}$ their Sylvester matrix. Then,
\begin{equation}
\label{eq:cgd-syl}
dim(\nullity(S_{f,g}) = \deg(\gcd(f, g)) \enspace .
\end{equation}
\begin{proof}
Notice that $S_{f,g}$ has size $m\times m$, where $m = \deg(f)+\deg(g)$. The idea is to compute the null space $\nullity(S_{f,g}^\top)=\{z \in \F_q^m : S_{f,g}^\top z^\top = \underbar{0}\}$ of the transposed Sylvester matrix. For any $z \in \nullity(S_{f,g}^\top)$ we write $z = (w \cat v)$ as the concatenation of the vectors $w \in \F_q^{\deg(g)}$ and $v \in \F_q^{\deg(f)}$. Next, we associate to $w$ and $v$ two polynomials $s,t$ respectively defined as:
\begin{align}
\label{eq:st}
    s(X) &= w_0 + w_1X + w_2X^2 + \cdots + w_{\deg(g)} X^{\deg(g)} \enspace , \\
    t(X) &= v_0 + v_1X + v_2X^2 + \cdots + v_{\deg(f)} X^{\deg(f)} \enspace .
\end{align}
where clearly $\deg(s) \le \deg(g)$ and $\deg(t) \le \deg(f)$. Then we have that $S_{f,g}^\top z$ can be written in polynomial form as:
\begin{equation}
\label{eq:bz}
f(X) s(X) + g(X) t(X) = \gcd(f, g)(X) \left(f_0(X) s(X) + g_0(X) t(X)\right) \enspace ,
\end{equation}
for suitable $f_0, g_0 \in \F_q[X]$ that are relatively prime. Therefore, $z$ belongs to the null space of $S_{f,g}^\top$ if and only if
\begin{equation}
\label{eq:bz-st}
f_0(X) s(X) + g_0(X) t(X) = 0 \enspace .
\end{equation}
By taking this identity modulo $g_0$, and omitting from now on the $(X)$ notation, we obtain
\begin{equation}
f_0 s \equiv 0 \pmod{g_0}  \enspace .
\end{equation}
Since $f_0$ and $g_0$ are coprime, we have $g_0 \mid s$, thus $s = g_0 p$ for some $p \in \F_q[X]$. Further, note that $\deg(p) = \deg(s) - \deg(g_0) \le \deg(g) - \deg(g_0) = \deg(\gcd(f,g))$. By replacing this in~\eqref{eq:bz-st} we get
\begin{equation}
\label{eq:bz-st-1}
f_0 g_0 p + g_0 t = g_0 (f_0 p + t) = 0 \enspace ,
\end{equation}
hence $t = - f_0 p$. Thus, $z$ belongs to the null space if and only if $(s, t)$ is of the form $(g_0 p, -f_0 p)$ for some $p$ with degree at most $\deg(\gcd(f,g))$. The dimension of the nullspace of (the transpose of) $S_{f,g}$ is then $\deg(\gcd(f,g))$. \qed
\end{proof}
\end{lemma}

We can now prove our main result: the minimum distance of a Grassmannian code $\mathcal{C}_\mathcal{F}$ generated by a family $\mathcal{F}$ of linear CA of diameter $d$ is determined by the largest degree of the pairwise GCD computed over the polynomials that define the local rules.
\begin{theorem}
\label{thm:min-dist}
Let $\mathcal{F}$ be a family of linear CA of length $2k$, each defined by a linear local rule of diameter $d$ where $k=d-1$. Then, the minimum distance of the Grassmannian code $\mathcal{C}_{\mathcal{F}}$ generated by $\mathcal{F}$ is equal to:
\begin{equation}
    \label{eq:min-dist-ca}
    D(\mathcal{C}_{\mathcal{F}}) = 2k - 2\cdot \max_{F, G \in \mathcal{F}} \left\{ \deg(\gcd(P_f,P_g)) \right\} \enspace ,
\end{equation}
where $P_f,P_g$ are the polynomials associated to the local rules of $F$ and $G$.
\begin{proof}
By Equation~\eqref{eq:dist-ker}, the distance between any two kernels $ker(F), ker(G)$ in $\mathcal{C}_\mathcal{F}$ is equal to $2k - 2 dim(ker(F) \cap ker(G))$. Hence, to determine $D(\mathcal{C}_{\mathcal{F}})$, we need to compute
\begin{equation}
\label{eq:max}
\max_{F, G \in \mathcal{F}} \left\{ dim(ker(F) \cap dim(ker(G)) \right\} \enspace .
\end{equation}
Recall that, by Remark~\ref{rem:int-ker}, the nullspace of the Sylvester matrix $M_H$ defined by $F$ and $G$ is the intersection of $ker(F)$ and $ker(G)$. Therefore, we have
\begin{equation}
    dim(ker(F) \cap dim(ker(G)) = \nullity(M_H) \enspace .
\end{equation}
Now, by Lemma~\ref{lm:rank-syl}, we have that $\nullity(M_H) = \deg(\gcd(f,g))$. We can thus rewrite~\eqref{eq:max} as:
\begin{equation}
\label{eq:max-1}
\max_{F, G \in \mathcal{F}} \left\{ dim(ker(F) \cap dim(ker(G)) \right\} =  \max_{F, G \in \mathcal{F}} \left\{ \deg(\gcd(f,g)) \right\} \enspace ,
\end{equation}
which proves our theorem. \qed
\end{proof}
\end{theorem}

\section{Equidistant Constant Dimension Codes from Linear CA}
\label{sec:cdc-ca}
In the previous section we proved that the minimum distance of a Grassmannian code generated by a family of linear CA depends on the maximum degree of the pairwise GCDs of their associated polynomials. We now analyze how large such a code can be by considering some specific cases.

For a given minimum distance $\delta$, one ideally wants to define a subspace code in such a way that it contains as many codewords as possible. To phrase it differently, we want to find the maximum number of degree $k$ polynomials in $\F_q[X]$, such that their pairwise GCD has degree at most $t = k - \delta/2$.

The optimal case of the highest minimum distance occurs when $t = 0$. As a matter of fact, this happens when all polynomials that define the linear CA in the family $\mathcal{F}$ are pairwise coprime, as shown in the next result:
\begin{lemma}
\label{lm:opt}
Let $\mathcal{C}_\mathcal{F}$ be a Grassmannian code generated by a set $\mathcal{F}$ of linear CA $F_1,\cdots,F_r: \F_q^{2k} \to \F_q^k$, defined by the local rules $f_1,\cdots, f_r: \F_q^d \to \F_q$ of diameter $d$ where $k=d-1$. Suppose that for each $F_i,F_j \in \mathcal{F}$ with $i\neq j$ the polynomials $P_{f_i}, P_{f_j}$ associated to the local rules respectively of $F_i$ and $F_j$ are coprime, that is $\gcd(P_{f_i}, P_{f_j}) = 1$. Then, the minimum distance of the code is:
\begin{equation}
\label{eq:opt-dist}
D(\mathcal{C}_\mathcal{F}) = 2k \enspace . 
\end{equation}
\end{lemma}
Notice that the code in Lemma~\ref{lm:opt} is also \emph{equidistant}: every pair of codewords in $\mathcal{C}_\mathcal{F}$ has distance $2k$. The maximum cardinality achievable by a subspace code of this kind corresponds to the size $N_k$ of the largest family of pairwise coprime polynomials with degree $k$ and nonzero constant term. This problem has already been addressed in~\cite{mariot20}, where the authors algorithmically build such sets of polynomials and prove their maximality. Specifically, $N_k$ is equal to:
\begin{equation}
\label{eq:char-max-mols}
N_k = I_k + \sum_{j=1}^{\lfloor \frac{k}{2} \rfloor} I_j \enspace ,
\end{equation}
where, for all $n \in \N$, $I_n$ is the cardinality of the set $\mathcal{I}_n$ of irreducible polynomials of degree $n$, which can be computed through \emph{Gauss's formula}~\cite{gauss-irr}:
\begin{equation}
    \label{eq:gauss_formula}
    I_n := |\mathcal{I}_n| = \frac{1}{n} \sum_{d | n} {\mu(d) q^{n / d}} \enspace ,
\end{equation}
with $\mu(\cdot)$ denoting the \emph{Möbius function}~\cite{lidl1997finite}.

If we relax the assumption on the minimum distance, allowing it for being non-optimal, we get into the generic case, where we allow the pairwise GCDs to have degree at most $t > 0$. In what follows, let us define the set of all monic polynomials of degree $k$ with nonzero constant term as:
\begin{equation}
\label{eq:polyfq}
    \text{Poly}_k(\F_q) := \left\{ f \in \F_q[X] : f \, \text{monic}, \, f(0) \ne 0, \, \deg(f) = k \right\} \enspace .
\end{equation}
Further, let $\text{CD}_{k,t}(\F_q)$ be the family of subsets of $\text{Poly}_k(\F_q)$ such that the degree of the pairwise GCDs is at most $t$:
\begin{equation}
    \text{CD}_{k,t}(\F_q) := \left\{ S \subseteq \text{Poly}_k(\F_q) : \forall f_1, f_2 \in S , \, \deg\left(\gcd\left(f_1, f_2\right)\right) \le t \right\} \enspace .
\end{equation}

The goal is to find a maximal element of $\text{CD}_{k,t}(\F_q)$ and its cardinality, that is $\max_{S \in \text{CD}_{k,t}(\F_q)}{|S|}$. This general case is quite tricky to handle. For this reason, here we address an intermediate problem, where we assume that \emph{all pairs of polynomials have exactly the same GCD} $g \in \F_q[X]$ \emph{with degree $t$}. This corresponds to finding the largest set in:
\begin{equation}
    \text{CF}_{k,g}(\F_q) := \left\{ S \subseteq \text{Poly}_k(\F_q) : \forall f_1, f_2 \in S , \, \gcd(f_1, f_2) = g \right\} \enspace .
\end{equation}
Remark that the resulting Grasmannian code is again equidistant in this case, with minimum distance $2k - 2t$.

The fixed polynomial $g$ is a common divisor of all the polynomials in the set $S$. So, for any polynomial $f \in \text{Poly}_k(\F_q)$, we can find $f'$ such that $f = gf'$, with $\deg(f') = k - t$.
To build our maximal set $S$ and compute its size, we can therefore use the same approach as in~\cite{mariot20} applied to $\text{Poly}_{k - t}(\F_q)$. In particular, we can build a set $S \in \text{CF}_{k,g}(\F_q)$ by adopting a straightforward variation of the algorithm {\sc Construction-Irreducible}. The modified pseudocode is reported below:

\begin{description}
\item[{\sc Construction-Uniform-GCD}$(k, g)$]
\item[Initialization:] Initialize set $T$ to $\mathcal{I}_{k - t}$, where $t = \deg(g)$
\item[Loop:] For all $1 \le i \le \left \lfloor \frac{k-t}{2} \right \rfloor$ do:
  \begin{enumerate}
    \item Build set $T_i$ by multiplying each polynomial in $\mathcal{I}_i$ with a distinct polynomial in $\mathcal{I}_{k-t-i}$
    \item Add set $T_i$ to $T$
  \end{enumerate}
\item[Final step:]
If $k-t$ is odd, build set $T_{(k-t-1)/2}$ by
multiplying each polynomial in $\mathcal{I}_{(k-t-1)/2}$ with a distinct
polynomial in $\mathcal{I}_{(k-t+1)/2}$, and add $T_{(k-t-1)/2}$ to
$T$.
If $k-t$ is even, build set $T_{(k-t)/2}$ by
squaring each irreducible polynomial in $\mathcal{I}_{(k-t)/2}$, and add
$T_{(k-t)/2}$ to $T$.
Finally, define the set $S := \left\{ g f' : f' \in T \right\}$.
\item[Output:] return $S$
\end{description}

It is easy to see that the set built by the above algorithm belongs to $\text{CF}_{k,g}(\F_q)$: every element of $S$ is monic since the product of monic polynomials, it has constant coefficient non-zero since both factors do as well, and it has degree $k$. Moreover, since the intermediate set $T$ belongs to $\text{CF}_{k,1}(\F_q)$ thanks to~\cite{mariot20}, it follows that for all $f'_1, f'_2 \in T$ we have $\gcd(f'_1, f'_2) = 1$ and thus $\gcd(g f'_1, g f'_2) = g$.

Therefore, by following the same arguments in~\cite{mariot20}, we can see that the cardinality of such set is:
\begin{equation}
    |S| = I_{k-t} + \sum_{i = 1}^{\left \lfloor \frac{k-t}{2} \right \rfloor}{I_i} \enspace .
\end{equation}

Finally, regarding the maximality, we pick a maximal element $A \in \text{CF}_{k,g}(\F_q)$ and define $A' := \left\{ f / g : f \in A \right\}$. Then, the proof can just follow the argument of~\cite{mariot20} by applying it to the set $A'$.

\section{Conclusions and Future Works}
\label{sec:outro}
In this paper, we started to investigate subspace codes generated by families of linear CA. We first remarked that the subspaces codes generated by CA with uniform diameter are Grassmannian. Then, we proved that the minimum distance of such codes is determined by the maximum degree of the pairwise GCDs of the polynomials associated to the local rules. Finally, we analyzed the maximal cardinality achievable by these subspace codes, considering two particular cases. The first one corresponds to the problem of counting how many pairwise coprime monic polynomials of fixed degree and nonzero constant term over a finite field exist, already addressed in~\cite{mariot20}, and we remarked that the resulting Grassmannian codes achieve the highest possible minimum distance $2k$. Next, we focused on the case where the polynomials have the same pairwise GCD. We presented a modified version of the algorithm in~\cite{mariot20} to construct such a set of polynomials, and we showed that it is maximal. 

There are several interesting directions to explore for future research. The most straightforward generalization would be to build Grassmannian codes from sets of linear CA where the underlying polynomials do not have the same pairwise GCD, but the degree is still fixed. The next step would then be to build and count the codes by setting an upper bound on the degree of the GCD. In this way, the cardinality of the optimal code can be determined exactly. Further, a comparison with the Grassmannian codes obtained with our method against those already published in the literature is in order, since the optimal case of our construction is a specific instance of the partial spreads codes introduced in~\cite{gorla14}. Finally, we would like to investigate the \emph{decoding} aspect of our subspace codes, and study if it is possible to exploit the parallel nature of the CA to build an efficient decoder. We believe that the inversion algorithm for mutually orthogonal CA presented in~\cite{mariot18a} represents a viable starting point to investigate this direction.

\bibliographystyle{abbrv}
\bibliography{bibliography}

\end{document}